# Non-invasive super-resolution imaging through dynamic scattering media

Dong Wang [1,2,#], Sujit K. Sahoo[1,3,#] and Cuong Dang[1,*]

[1] Centre for Optoelectronics and Biophotonics (COEB), School of Electrical and Electronic Engineering, The Photonics Institute (TPI), Nanyang Technological University Singapore, 50 Nanyang Avenue, 639798, Singapore

[2] Key Laboratory of Advanced Transducers and Intelligent Control System, Ministry of Education, and Shanxi Province, College of Physics and Optoelectronics, Taiyuan University of Technology, Taiyuan 030024, China

[3] School of Electrical Sciences, Indian Institute of Technology Goa, Goa 403401, India

[#] Dong Wang and Sujit K. Sahoo contribute equally.

*Corresponding author, E-mail: HCDang@ntu.edu.sg

## Abstract

Super-resolution imaging with advanced optical systems has been revolutionizing technical analysis in various fields from biological to physical sciences. However, many objects are hidden by strongly scattering media such as rough wall corners or biological tissues that scramble light paths, create speckle **patterns and hinder object's visualization, let alone super-resolution imaging. Here, we realize a method to do non-invasive super-resolution imaging through scattering media based on stochastic optical scattering localization imaging (SOSLI) technique. Simply by capturing multiple speckle patterns of photo-switchable emitters in our demonstration, the stochastic approach utilizes the speckle correlation properties of scattering media to retrieve an image with more than five-fold resolution enhancement compared to the diffraction limit, while posing no fundamental limit in achieving higher spatial resolution. More importantly, we demonstrate our SOSLI to do non-invasive super-resolution imaging through not only optical diffusers, i.e. static scattering media, but also biological tissues, i.e. dynamic scattering media with decorrelation of up to 80%. Our approach paves the way to non-invasively visualize various samples behind scattering media at unprecedented levels of detail.**

## Introduction

Optical imaging beyond the diffraction-limit resolution has enabled incredible tools to advance science and technology from non-invasive investigation of the interior of biological cells[1,2] to chemical reactions at single molecule levels[3]. Super-resolution stimulated emission depletion (STED) microscopy[4] has been progressed rapidly to achieve three-dimensional (3D) imaging with super-high spatiotemporal precision[5,6]. For single-molecule detection and localization approaches[7,8], such as stochastic optical reconstruction microscopy (STORM) or photo-activated localization microscopy (PALM), positions of photo-switchable probes are determined as centers of diffraction-limited spots. Repeating multiple imaging processes, each with a stochastically different subset of active fluorophores, allows positioning a number of emitters at few-nanometer resolution so that a high-resolution image is reconstructed[9]. After these pioneering techniques, the field of super-resolution microscopy has developed rapidly with various other techniques[10-12] to bring the optical microscopy within the resolution of electron microscopy. However, the requirement of sample transparency makes the super-resolution microscopy techniques impossible to access objects, which are hidden by strongly scattering media (Fig. 1a and supplementary Fig. S1, S2) such as biological tissues, frosted glass or around rough wall corners. These media do not absorb light significantly; however, they create noise-like speckle patterns[13] and challenge even our low-resolution visualization of samples.

Many approaches have been demonstrated to overcome the scattering effects and enabled imaging or focusing capability through scattering media. The most straightforward approaches utilize ballistic photons[14], such as optical coherence tomography[15] or multi-photon microscopy[16]. However, strongly scattering media significantly reduce number of ballistic photons and lower the signal tremendously[17]. Some techniques require a guide star or an access on the other side of scattering media to characterize or reverse their scattering effects before imaging such as wave-front shaping techniques[18-21], transfer matrix measurement[22,23]. The memory effect of light through scattering media[24,25] implying a shift-invariant point spreading function (PSF), allows imaging by deconvolution[26-28] of a speckle pattern with the PSF, which is measured invasively. A scattering medium with a known PSF is a scattering lens that, in turn, is a low pass filter like any conventional lens. Deconvolution recovers and enhances the transmission of high frequency components (i.e. a sharper cut-off low-pass filter) therefore, resolution is slightly higher than the diffraction limit (Fig. 1b and supplementary Fig. S2d-f). Deconvolution imaging currently provides the best-resolution images from speckle patterns. Each measured PSF is only valid for one scattering medium, therefore, cannot be used for dynamic scattering media. Non-invasive imaging through scattering media where the image is retrieved without characterizing scattering media is desired for real applications. Diffuse optical tomography[29,30] and time-of-flight imaging[14,29,31] are capable of seeing though scattering layers and around corners non-invasively; however, the spatial resolution is much lower than the optical diffraction limit. Thanks to the shift-invariance speckle-type PSF of thin scattering media, the 2D image and even the 3D image of a sample can be revealed non-invasively from the speckle patterns by a phase retrieval algorithm[32-34]. The limited performance of the algorithm and the camera, together with presence of noise and sample's complexity, usually makes the image retrieval process fail or converge with some artifacts and slightly lower resolution compared to the diffraction limit (Fig. 1c and supplementary Fig. S2g-i).

Here, we present our stochastic optical scattering localization imaging (SOSLI) technique to do non-invasive super-resolution imaging through scattering media. The technique only requires an imaging sensor to capture speckle patterns created by blinking emitters behind a scattering medium; no other optics or complicated alignment are needed (Fig. 1a, Fig. 2a, and supplementary Fig. S1). The positions of emitters in each stochastic frame are first determined at very high precision from the corresponding speckle pattern, and then a super-resolution image of the full sample is reconstructed by superposing a series of emitter position images (Fig. 1d). We demonstrate the image reconstruction with resolution beyond the diffraction limit by a factor of five as a proof of concept, while there is no fundamental limit for SOSLI on resolution, similar to current super-resolution microscopy techniques. More interestingly, the localization algorithm is based on a single-shot speckle pattern with minimum correlation among adjacent patterns; therefore, we develop adaptive SOSLI to do super-resolution imaging through dynamic scattering media such as fresh chicken eggshell membrane with decorrelation of up to 80%. Our SOSLI demonstrates a desired technique to see through translucent media such as biological tissues or frosted glass with unprecedented clarity.

## Stochastic Optical Scattering Localization Imaging (SOSLI)

An object $O$ consists of stochastically blinking emitters: $O = \sum_{i=1}^{N} O_i$, where $O_i$ is the $i^{\text{th}}$ blinking pattern (a subset of $O$) and $N$ is the total number of the blinking patterns. After light propagating through scattering media, each $O_i$ produces a speckle pattern $I_i$, captured by a camera (Fig. 2a). If object size is within the memory effect of the scattering media, the PSF is shift-invariant and speckle-type, therefore the speckle pattern $I_i$ (Fig. 2b) of object $O_i$ preserves the object's autocorrelation (Fig. 2c). The image of $O_i$ can be retrieved from its autocorrelation by an iterative phase retrieval algorithm[33] (Fig. 2d). The limitation in the camera's bit depth, photon budget and performance of phase retrieval algorithms in presence of image acquisition noise degrade the diffraction-limit resolution of this non-invasive retrieval image. However, a standard localization algorithm[35,36] is employed to find the position of emitters at very high resolution (Fig. 2e) and remove algorithm artifacts. Similar to other localization microscopy techniques, the precision is higher for spatially sparse emitter samples where only one emitter is temporally active in a diffraction-limited region. The sharp and clear image $O'_i$ presents the precise relative emitter positions of pattern $O_i$, while losing their exact positions because $O'_i$ is only retrieved from autocorrelation of $O_i$ through autocorrelation of $I_i$. The estimated PSF of the scattering medium can be retrieved by deconvolution (Fig. 2f): $PSF' = Deconvol(I_i, O'_i)$, which is also shifted in comparison to the actual PSF because of losing exact emitter positions in $O'_i$.

Next, a series of clean super-resolution images $O'_j$ with emitter positions is reconstructed for a corresponding series of stochastic patterns $O_j$ by deconvolution of its corresponding speckle pattern $I_j$ with the estimated $PSF'$ and localization as presented in Fig. 2g. A super-resolution image of the full sample (Fig. 2h) is now reconstructed by superposing all individual images as: $O' = \sum_{j=1}^{N} O'_j$, which represents object $O$ with an arbitrary position. This principle is valid provided that PSF does not change among the group of $I_j$. For comparison, we present the typical simulation image (Fig. 2i) retrieved from autocorrelation of a single speckle pattern, in which simulation parameters are similar with the exception that all the emitters are on. The simulated diffraction limit is about 3.2 pixels. This current state-of-the-art technique for non-invasive imaging through scattering media shows a blurry image where the low spatial frequency presents the diffraction limit of the optical system together with some artifact from the phase retrieval algorithm. In contrast, the image reconstructed by SOSLI is much sharper (Fig. 2h).

## Super-resolution imaging through a ground glass diffuser

To prove our concept, we first demonstrate SOSLI for non-invasive super-resolution imaging through a ground glass diffuser. Microscopic objects comprising multiple stochastic blinking emitters are created by de-magnifying projector images through a microscope objective. The de-magnifying image of each pixel in a digital micro-mirror device (DMD) is an intermittent emitter with a size of about 1.34 µm (Supplementary Fig. S1a). The microscopic object is placed 10 mm behind the ground glass diffuser, which is kept unknown in all demonstrations. The incoherent light from the object propagating through the optical diffuser is recorded by a monochromatic camera, which is 100 mm in front of the diffuser. An iris with diameter of 1 mm is placed immediately after the optical diffuser to act as the aperture of the imaging system. A larger iris size enhances the diffraction limit of the imaging system and achieve a sharper image (supplementary Fig. S2); however, it reduces the speckle contrast that is vital for phase retrieval approach.

Figure 3 shows the experimental results for three different imaging approaches.

Single-shot non-invasive imaging through scattering media is performed in Fig. 3a, where all emitters are on. The result is recovered from the autocorrelation of a single speckle pattern by the phase retrieval algorithm. The image is very blurred, and we cannot distinguish 2 lines with a gap of 4 µm between them (Fig. 3b). We can calculate the diffraction limit of our system as $0.61\,\lambda/\text{NA} = 6.7\,\mu m$, where $\lambda = 550$ nm and NA = 0.05. Beside the diffraction limit, the performance of the phase retrieval algorithm in the presence of experimental noise also degrades the image quality and limits the resolution. With the DMD projector, we can measure the PSF by turning on a single pixel at the center only and capture its speckle pattern. Such an "invasive guiding star" for the PSF measurement allows us to calculate the image by deconvolution and significantly enhances the resolution (Fig. 3c). The invasive deconvolution approach is more deterministic, robust to the noise and enhances the high spatial frequency components of the image. This allows us to distinguish the 2 lines with a 4-µm gap between them (Fig. 3d). However, we still cannot see the gap of 2.68 µm between 2 lines. Most strikingly, our super-resolution image reconstructed non-invasively by SOSLI is remarkably clear as presented in Fig. 3e. We can resolve very well all the smallest features of our sample, i.e. 2 thin lines (1.34 µm width) with a gap of 1.34 µm in between (Fig. 3f). The smallest sample feature is smaller than the diffraction limit by a factor of 5. Figure 3e-f also clearly illustrates that the capability of SOSLI is far beyond our sample's smallest features (pixel size), which are currently limited by the projector and optics of the sample creating system.

It is worth to highlight some important factors in our SOSLI performance. Supplementary Fig. S3 presents more detail for the reconstruction process in which the localization is important to remove all the background noise and artifacts in the emitter image resulting from the phase retrieval algorithm. This leads to a better estimation of PSF for a series of deconvolution calculation after that. In addition, the localization process also allows SOSLI to tolerate more errors in deconvolution process due to imperfect PSF estimation and noise in image acquisition (Supplementary Fig. S4-5). Our approach relies on stochastic emitter patterns to reconstruct a full object; therefore, the image quality is improved with more stochastic patterns (supplementary Fig. S6). Figure 4 presents some images of more complex objects for performance comparison among the three techniques. Similar to Fig. 3, the complex objects are best resolved with our SOSLI approach (Fig. 4a-c), while the retrieval image from autocorrelation of a single speckle pattern shows the poorest performance that also has some artifacts (Fig. 4d-f). The invasive imaging approach by deconvolution shows moderate performance in Fig. 4g-i. Obviously, our SOSLI for non-invasive imaging through scattering media goes far beyond the diffraction limit and surpasses all the current state-of-the-art imaging through scattering media, including both invasive and non-invasive techniques.

## Super-resolution imaging through a biological tissue

Our SOSLI demonstrations in Fig. 3 and Fig. 4 rely on a fixed PSF for reconstruction of multiple stochastic emitter patterns; and therefore, we cannot directly use for dynamic scattering media such as biological tissues. Figure 5a shows the decorrelation behaviors of PSFs for two different scattering media. For static scattering media such as ground glass diffusers, the PSF is a constant pattern and the correlation of 1 is achieved for any measured PSFs at any time. On the other hand, dynamic scattering media such as fresh chicken eggshell membrane, the PSF is gradually changed and the correlation with the initial one decreases with time. In our experiment for fresh chicken eggshell membrane, the correlation reduces from 1.0 to 0.2 after 300 measurements, with the fastest decay rate in the first 70 measurements (correlation decreases to 0.54). The reconstruction by SOSLI with a single estimated PSF shows a noisy and blurred image due to this decorrelation of the membrane (Fig. 5b). Supplementary Fig. S7 presents the deconvolution images from stochastic speckle patterns with an estimated PSF from the first speckle pattern. Obviously, the assumption of a static PSF in SOSLI does not hold in this case. For very large PSF decorrelations, the localization process cannot distinguish the emitters from strong background and artifacts in the deconvolution images, leading to the result in Fig. 5b.

We introduce an adaptive approach to demonstrate our SOSLI for super-resolution imaging through dynamic scattering media. We now utilize SOSLI to localize and then superpose emitters in 50 stochastic patterns, in which the fresh chicken eggshell membrane still can retain its PSF correlation of more than 60%. With a total collection of 300 stochastic patterns, we divide this into 11 sections, each containing 50 speckle patterns in which the first 25 patterns are overlapped with the previous section; the other 25 patterns are then overlapped with the next section (Supplementary Fig. S8). We can directly apply SOSLI with single estimated PSFs to reconstruct 11 super-resolution images, each one representing a part of the object. However, we cannot directly superpose these 11 images to reconstruct the full image because their absolute positions are lost in each SOSLI procedure. In fact, the relative positions among the images are the relative positions among their respective retrieved PSFs, which are achieved independently (similar to the discussion on estimated PSFs from different speckle patterns in supplementary Fig. S4, S5). The correlation between two retrieved PSFs is a bright spot with background; the intensity of the bright spot presents the correlation between two PSFs; and the position of the bright spot (relative to the center) presents the relative position between two PSFs. Because PSF correlation only reduces to about 65% between two adjacent sections in our experiment, the brightest spot in correlation between two estimated PSFs is very clear and its center is very easy to locate (similar to emitter localization). With this procedure, the relative positions of 11 images are identified (Supplementary Fig. S9). We only need to shift our individual super-resolution images to make the relative position zero before superposing them to receive a full image. Figure 5c presents the superposing result after alignment of their individual images. The final image with adaptive SOSLI is super-resolution, much clearer and less noise compared to SOSLI with a static PSF (Fig. 5b). For comparison, the low-resolution images obtained by the phase retrieval algorithm and invasive deconvolution of this object through chicken eggshell membrane are similar to Fig. 1b-c, respectively. By doing adaptive SOSLI, we reconstruct a super-resolution image non-invasively through dynamic scattering media with effective correlation of more than 0.6 (shading area in Fig. 5a) while the actual correlation reduces to 0.2 during image acquisition. The procedure can continue with more stochastic pattern acquisition and the membrane is even completely decorrelated but the effective correlation for adaptive SOSLI still can maintain at more than 60%. Our SOSLI with an adaptive PSF shows a very practical approach to do non-invasive super-resolution imaging through dynamic scattering media.

## Discussion and Conclusion

Our SOSLI relies on a shift-invariant speckle-type PSF of scattering media, which is valid if the hidden object behind the scattering media is within the memory effect region of the scattering media. Therefore, we can image a larger object with thinner scattering media. However, there is no fundamental physics to limit our SOSLI's resolution, similar to conventional super-resolution microscopy. In fact, we have practical challenges for SOSLI to reach high resolution. The most challenging requirement is the photon budget for each intermittent emitter, i.e. the number of photons per emitter per blink. For the localization approach in super-resolution microscopy[7-9] where all the photons go into the diffraction limit spots, the resolution can be enhanced by $\sqrt{N}$ where $N$ is the number of captured photons. With scattering media, we need to capture many more photons, because the photons are now scattered everywhere, form speckles, and we need multiple speckles to retrieve an image. The noise and bit depth of camera, as well as the sparsity of active emitters are also important factors that affect our computational approach, limiting the resolution.

It is important for any imaging technology to beat the dynamics of both the object and the environment. Like any other localization approach[7-9], our SOSLI requires a static object during the whole image acquisition process. However, the adaptive SOSLI can significantly mitigate the environment dynamics (i.e. the decorrelation of scattering media). It is worth to note that our method for localizing emitters in a stochastic speckle pattern relies on a single-shot image, and then the final super-resolution image relies on alignment of the localized-emitter patterns. The former implies that a stochastic speckle pattern can be captured sufficiently fast to beat the dynamics of the scattering media. The latter is more important for SOSLI. In extremely dynamic scattering media, the deconvolution might fail even when using the estimated PSF for immediately next speckle pattern because of fast decorrelation. We can then apply the most adaptive SOSLI by retrieving every emitter pattern from its speckle autocorrelation by phase retrieval algorithm then localization. The only requirement for the most adaptive approach is that the scattering media do not decorrelate completely between two consecutive shots. In our demonstration (supplementary Fig. S10), only 20% correlation is sufficient to determine the relative positions between two super-resolution images. Superposing then can be carried out after proper alignment to achieve a super-resolution image.

In summary, we have presented our simulation and proof-of-concept demonstration of SOSLI for non-invasive super-resolution imaging through both static and dynamic scattering media. We only need a camera to capture multiple images of scattered light from stochastic emitters behind scattering media, and then our computational approach will localize these emitters non-invasively to reconstruct a super-resolution image. Our experimental results show that SOSLI enhances resolution by a factor of 5 compared to the diffraction limit, showing features with considerably more detail compared to both state-of-the-art invasive and non-invasive imaging through scattering media. The demonstrated resolution enhancement is currently limited by our sample preparation while the SOSLI technique presents no fundamental limit to achieve higher resolution. The adaptive SOSLI allows super-resolution imaging non-invasively through highly dynamic scattering media with decorrelation of up to 80% while capturing two consecutive speckle patterns. Our SOSLI demonstration shows a promising approach for optical imaging through dynamic turbid media, such as biological tissue, with unprecedented clarity.

## Method

**Scale bar:** All the experimental results of recovered images show a scale bar of 10 camera pixels that is equivalent to 65 µm in the imaging plane. This is corresponding to 6.5 µm on the object plane because the magnification is 10. However, we do not know the scale bar on the object plane or magnification of the imaging system in non-invasive approach because the distance from the object to scattering media is unknown. We can only resolve the sample by angle resolution. The scale bar of 65 µm in the imaging plane is equivalent to the angle of 0.65 mrad.

**Data processing:** In all experiments, the resolution of the raw camera images is 2560×2160

pixels. We crop them into a resolution of 2048×2048 pixels for implementations of all the mentioned techniques in this work. The final reconstructed images are cropped to a square window with dimensions ranging between 75×75 pixels and 150×150 pixels (depending on the imaged object dimensions). Algorithms are developed in Matlab and run on a normal PC (Intel Core i7, 16 GB memory). A typical procedure for SOSLI with 300 speckle patterns takes 2-3 minutes.

## Funding

The research is financially supported by Singapore Ministry of Health's National Medical Research Council (NMRC): CBRG-NIG (NMRC/BNIG/2039/2015), Ministry of Education – Singapore (MOE): MOE-AcRF Tier-1 (MOE2017-T1-002-142), and Nanyang Technological University (NTU).

## Acknowledgements

Authors specially thank Professor Sylvain Gigan and his research group at Laboratoire Kastler Brossel, Sorbonne Université, École Normale Supérieure–Paris Sciences et Lettres (PSL) Research University, Paris, France for great discussions and providing suggestions for improvement. We would like to thank Xiangwen Zhu, Vinh Tran, Dr. Dayan Li, Dr. Dongliang Tang, and Dr. Huy Lam at NTU Singapore for fruitful discussions and useful feedbacks. We would like to thank Dr. Philip Anthony Surman for proof reading.

# Author contributions

S.K.S. performed the numerical simulations. C.D., D.W. and S.K.S. designed the initial experiments. D.W. performed the experiments with S.K.S.'s participation. All authors discussed, analyzed and took responsibility for the results and content of the paper. C.D. and D.W. wrote the manuscript with S.K.S.'s contributions. C.D. supervised and contributed to all aspects of research.

# References

1   Schermelleh, L. *et al.* Super-resolution microscopy demystified. *Nature Cell Biology* **21**, 72-84, (2019).

2   Neef, J. *et al.* Quantitative optical nanophysiology of Ca2+ signaling at inner hair cell active zones. *Nature Communications* **9**, 290, (2018).

3   Pujals, S., Feiner-Gracia, N., Delcanale, P., Voets, I. & Albertazzi, L. Super-resolution microscopy as a powerful tool to study complex synthetic materials. *Nature Reviews Chemistry* **3**, 68-84, (2019).

4   Hell, S. W. & Wichmann, J. Breaking the diffraction resolution limit by stimulated emission: stimulated-emission-depletion fluorescence microscopy. *Opt Lett* **19**, 780-782, (1994).

5   Vicidomini, G., Bianchini, P. & Diaspro, A. STED super-resolved microscopy. *Nature Methods* **15**, 173, (2018).

6   Hell, S. W. Toward fluorescence nanoscopy. *Nature Biotechnology* **21**, 1347-1355, (2003).

7   Betzig, E. Proposed method for molecular optical imaging. *Opt Lett* **20**, 237-239, (1995).

8   Betzig, E. *et al.* Imaging Intracellular Fluorescent Proteins at Nanometer Resolution. *Science* **313**, 1642-1645, (2006).

9   Balzarotti, F. *et al.* Nanometer resolution imaging and tracking of fluorescent molecules with minimal photon fluxes. *Science* **355**, 606-612, (2017).

10  Turcotte, R. *et al.* Dynamic super-resolution structured illumination imaging in the living brain. *Proceedings of the National Academy of Sciences* **116**, 9586-9591, (2019).

11  Tenne, R. *et al.* Super-resolution enhancement by quantum image scanning microscopy. *Nat Photonics* **13**, 116-122, (2019).

12  Kaldewey, T. *et al.* Far-field nanoscopy on a semiconductor quantum dot via a rapid-adiabatic-passage-based switch. *Nat Photonics* **12**, 68-72, (2018).

13  Goodman, J. W. *Speckle Phenomena in Optics: Theory and Applications*. (Roberts & Company, 2007).

14  Kang, S. *et al.* Imaging deep within a scattering medium using collective accumulation of single-scattered waves. *Nat Photonics* **9**, 253, (2015).

15  Siddiqui, M. *et al.* High-speed optical coherence tomography by circular interferometric ranging. *Nat Photonics* **12**, 111-116, (2018).

16  Hoover, E. E. & Squier, J. A. Advances in multiphoton microscopy technology. *Nat Photonics* **7**, 93, (2013).

17  Ntziachristos, V. Going deeper than microscopy: the optical imaging frontier in biology. *Nature Methods* **7**, 603, (2010).

18  Čižmár, T., Mazilu, M. & Dholakia, K. In situ wavefront correction and its application to micromanipulation. *Nat Photonics* **4**, 388, (2010).

19  Horstmeyer, R., Ruan, H. & Yang, C. Guidestar-assisted wavefront-shaping methods for focusing light into biological tissue. *Nat Photonics* **9**, 563, (2015).

20  Judkewitz, B., Wang, Y. M., Horstmeyer, R., Mathy, A. & Yang, C. Speckle-scale focusing in the diffusive regime with time reversal of variance-encoded light (TROVE). *Nat Photonics* **7**, 300, (2013).

21  Si, K., Fiolka, R. & Cui, M. Fluorescence imaging beyond the ballistic regime by ultrasound-pulse-guided digital phase conjugation. *Nat Photonics* **6**, 657, (2012).

22  Choi, Y. *et al.* Overcoming the Diffraction Limit Using Multiple Light Scattering in a Highly Disordered Medium. *Physical Review Letters* **107**, 023902, (2011).

23  Chaigne, T. *et al.* Controlling light in scattering media non-invasively using the photoacoustic transmission matrix. *Nat Photonics* **8**, 58, (2013).

24  Osnabrugge, G., Horstmeyer, R., Papadopoulos, I. N., Judkewitz, B. & Vellekoop, I. M. Generalized optical memory effect. *Optica* **4**, 886-892, (2017).

25  Freund, I., Rosenbluh, M. & Feng, S. Memory Effects in Propagation of Optical

Waves through Disordered Media. *Physical Review Letters* **61**, 2328-2331, (1988).

26   Tang, D., Sahoo, S. K., Tran, V. & Dang, C. Single-shot large field of view imaging with scattering media by spatial demultiplexing. *Applied Optics* **57**, 7533-7538, (2018).

27   Sahoo, S. K., Tang, D. & Dang, C. Single-shot multispectral imaging with a monochromatic camera. *Optica* **4**, 1209-1213, (2017).

28   Edrei, E. & Scarcelli, G. Memory-effect based deconvolution microscopy for super-resolution imaging through scattering media. *Scientific reports* **6**, 33558, (2016).

29   Lyons, A. *et al.* Computational time-of-flight diffuse optical tomography. *Nat Photonics*, (2019).

30   Eggebrecht, A. T. *et al.* Mapping distributed brain function and networks with diffuse optical tomography. *Nat Photonics* **8**, 448, (2014).

31   O'Toole, M., Lindell, D. B. & Wetzstein, G. Confocal non-line-of-sight imaging based on the light-cone transform. *Nature* **555**, 338, (2018).

32   Bertolotti, J. *et al.* Non-invasive imaging through opaque scattering layers. *Nature* **491**, 232-234, (2012).

33   Katz, O., Heidmann, P., Fink, M. & Gigan, S. Non-invasive single-shot imaging through scattering layers and around corners via speckle correlations. *Nat Photon* **8**, 784-790, (2014).

34   Okamoto, Y., Horisaki, R. & Tanida, J. Noninvasive three-dimensional imaging through scattering media by three-dimensional speckle correlation. *Opt Lett* **44**, 2526-2529, (2019).

35   Huang, B., Wang, W., Bates, M. & Zhuang, X. Three-Dimensional Super-Resolution Imaging by Stochastic Optical Reconstruction Microscopy. *Science* **319**, 810-813, (2008).

36   Rust, M. J., Bates, M. & Zhuang, X. Sub-diffraction-limit imaging by stochastic optical reconstruction microscopy (STORM). *Nature Methods* **3**, 793-796, (2006).

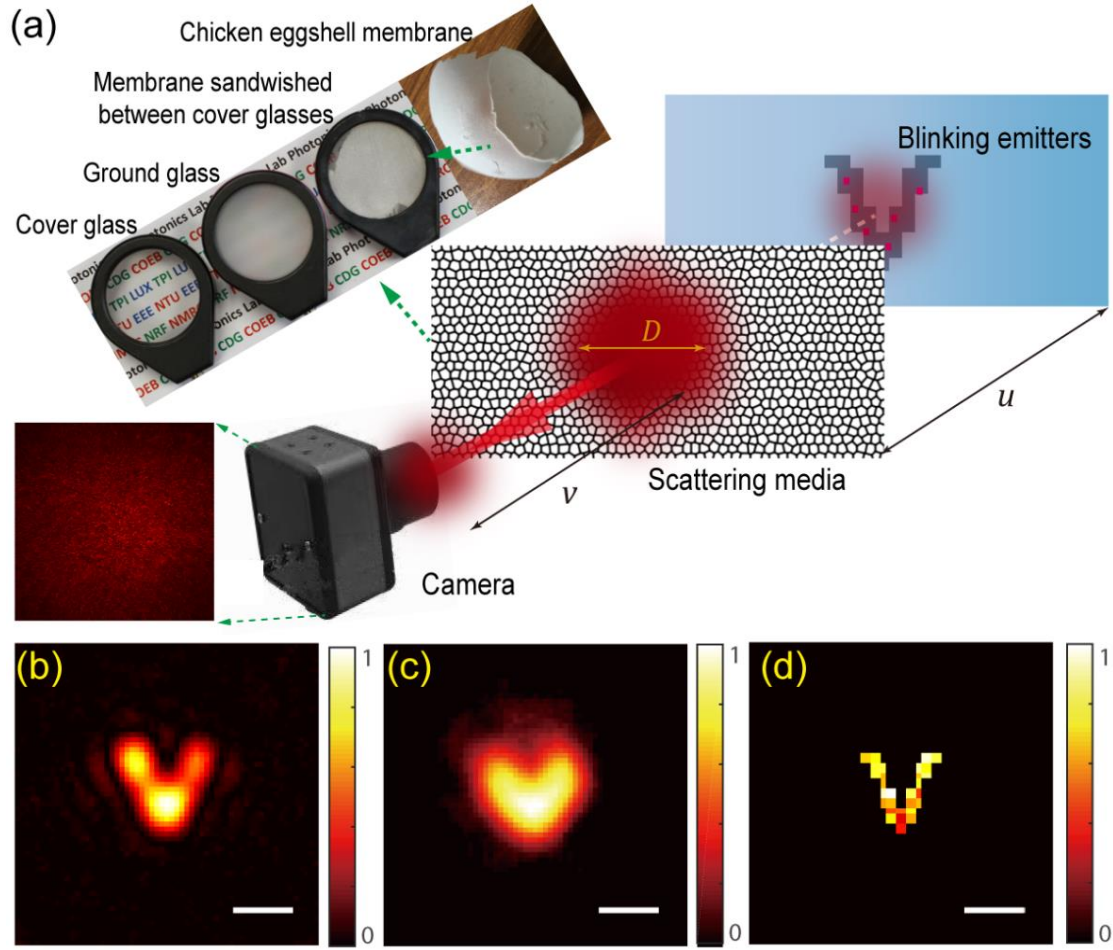

**Figure 1. Super-resolution imaging through scattering media with SOSLI in comparison to other imaging techniques.** (**a**) Schematic of SOSLI where incoherent light from blinking emitters hidden behind various scattering media is scattered and then captured by a camera. (**b & c**) Experimental demonstrations of the current state-of-art invasive and non-invasive imaging, respectively, through scattering media in the identical experimental setup. (**d**) A simulation demonstration of super-resolution imaging reconstructed by SOSLI. Simulation parameters are taken from the experiment in (b) and (c). Scale bars: 10 camera pixels, i.e. 65 µm on the imaging plane (see method).

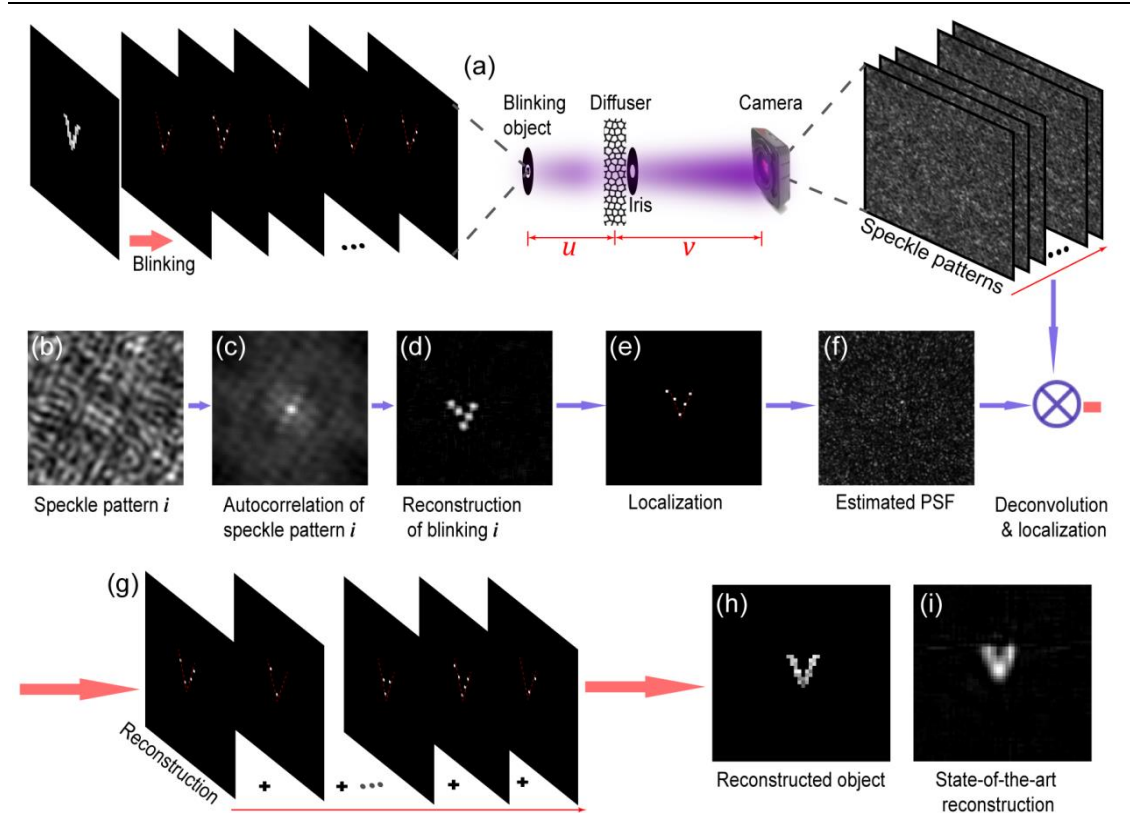

**Figure 2. Principle and simulation results of SOSLI.** (**a**) Object constitutes many intermittent emitters behind an optical diffuser; the iris defines the optical aperture of the imaging system and the camera captures speckle patterns. (**b**) A small portion of a typical speckle pattern. (**c**) Autocorrelation of the speckle pattern is similar to that of the emitter pattern. (**d**) A retrieved image from its autocorrelation. (**e**) Localized emitters from the retrieved image. (**f**) Estimated PSF' from the localized emitter image (e) and its corresponding speckle pattern (b). (**g**) A series of localized emitter images by deconvolution of the speckle patterns with the estimated PSF'. (**h**) A reconstructed image with a sub-diffraction-limit resolution by superposing all the individual localized emitter images. (**i**) A retrieved image from a single-shot speckle pattern when all emitters are on, i.e. the current state-of-art non-invasive imaging scheme[33].

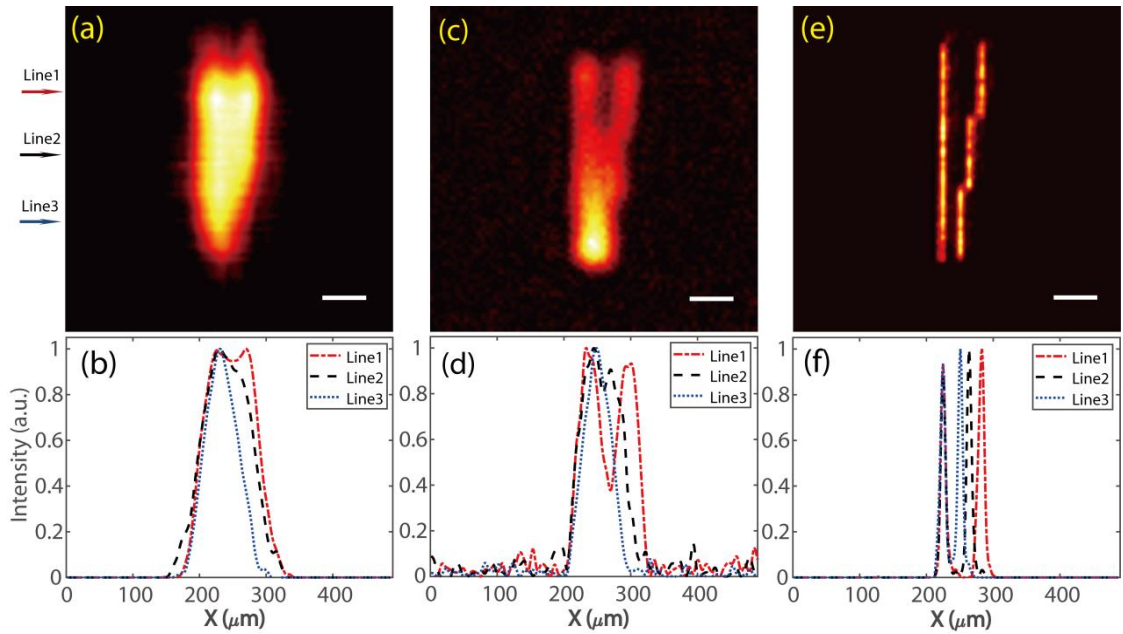

**Figure 3. Experimental results of imaging through a ground glass diffuser with different techniques. (a-b)** Single-shot non-invasive imaging retrieved from the speckle autocorrelation and its intensity profile, respectively. **(c-d)** Invasive imaging by deconvolution of the single speckle pattern with an invasively measured PSF and its intensity profile, respectively. **(e-f)** Non-invasive super-resolution imaging by our SOSLI and its intensity profile, respectively. Three arrows on the left indicate the three lines for cross-sectional intensity curves in figure b, d, f. Scale bars are 10 camera pixels, i.e. 65 µm on the imaging plane, and the values for X axis in graphs are on the imaging plane (see method).

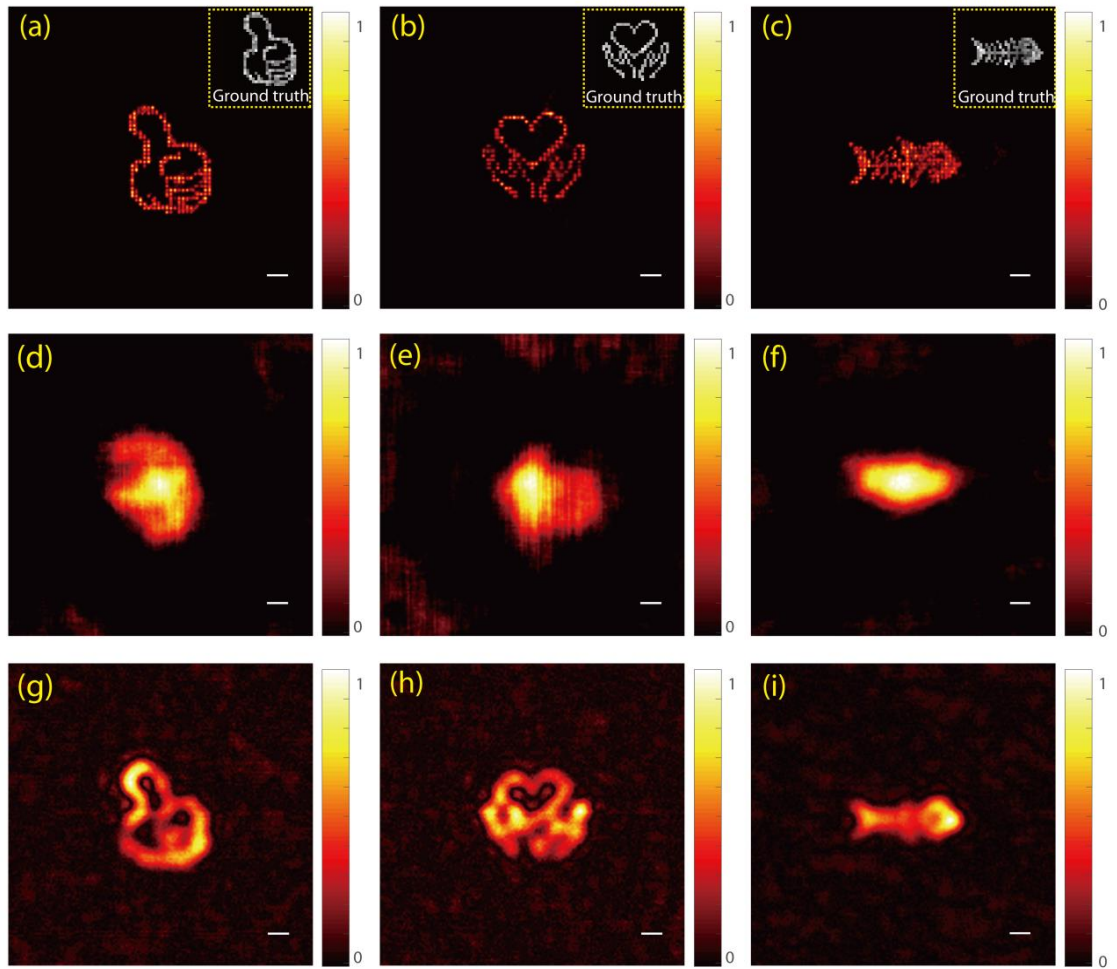

**Figure 4. Experimental demonstration of three techniques for imaging several complex objects hidden behind a ground glass diffuser.** **(a-c)** Our SOSLI approach for non-invasive super-resolution imaging. Insets are ground truth objects. **(d-f)** Non-invasive imaging retrieved from autocorrelation of a single speckle pattern for the ground truth samples in the insets of a, b and c, respectively. **(g-i)** Invasive imaging with an invasively measured PSF and deconvolution approach for the ground truth samples in the insets of a, b and c, respectively. Scale bars: 10 camera pixels, i.e. 65 µm on the imaging plane (see method).

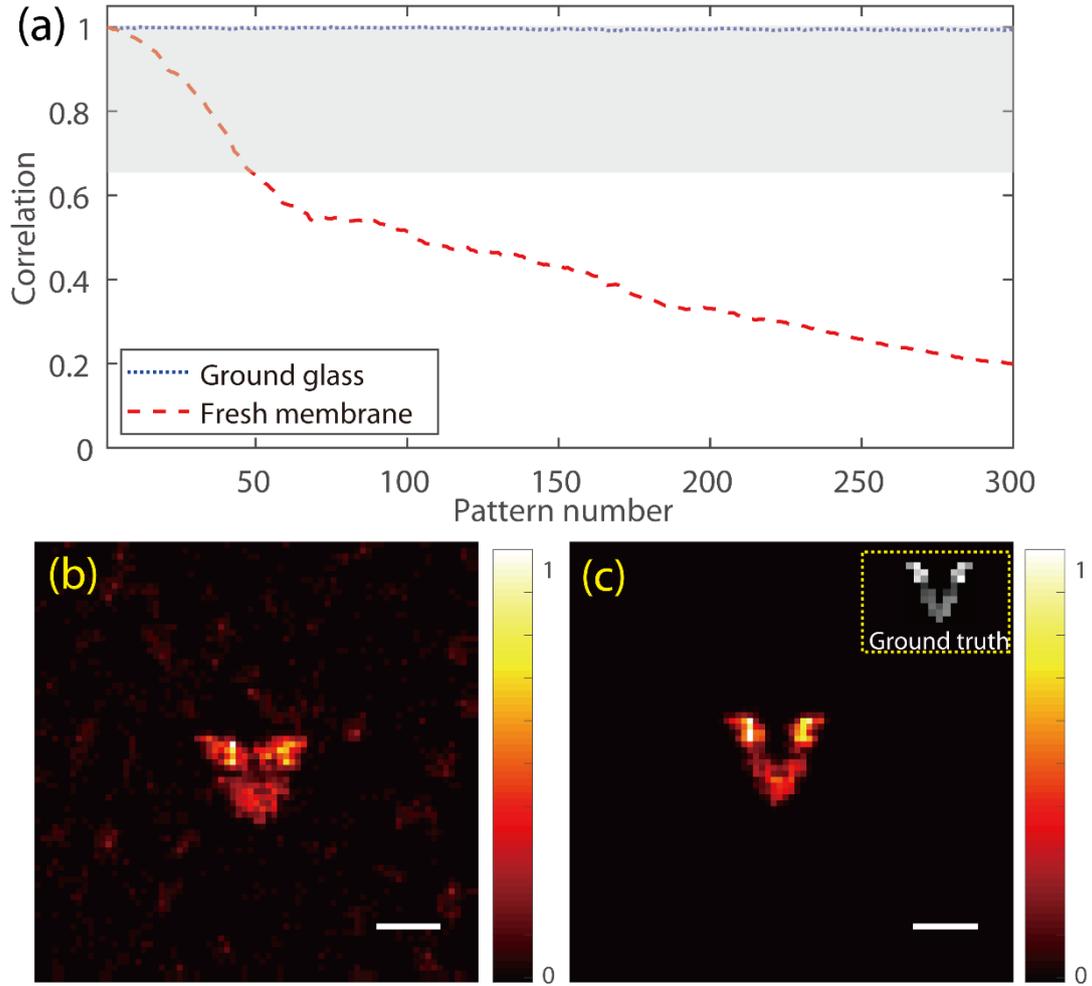

**Figure 5. Experimental demonstration of non-invasive super-resolution imaging through a fresh chicken eggshell membrane by SOSLI**. **(a)** Speckle correlation of PSFs at different measurement time for the static scattering medium (ground glass) and the dynamic one (fresh chicken eggshell membrane). **(b)** The reconstructed image by SOSLI with a single estimated PSF. **(c)** The reconstructed image by SOSLI with adaptive PSF estimation. Scale bars: 10 camera pixels, i.e. 65 µm on the imaging plane (see method).

# Supplementary material:
# Non-invasive super-resolution imaging through dynamic scattering media

Dong Wang[1,2,#], Sujit K. Sahoo[1,3,#] and Cuong Dang[1,*]

[1] Centre for Optoelectronics and Biophotonics (COEB), School of Electrical and Electronic Engineering, Nanyang Technological University Singapore, 50 Nanyang Avenue, 639798, Singapore

[2] Key Laboratory of Advanced Transducers and Intelligent Control System, Ministry of Education, and Shanxi Province, College of Physics and Optoelectronics, Taiyuan University of Technology, Taiyuan 030024, China

[3] School of Electrical Sciences, Indian Institute of Technology Goa, Goa 403401, India

[#] Dong Wang and Sujit K. Sahoo contribute equally.

[*] Corresponding author, E-mail: HCDang@ntu.edu.sg

## Optical experiment setup

The optical setup for experimental demonstration of stochastic optical scattering localization imaging (SOSLI) is depicted schematically in Supplementary Fig. S1. It consists of two parts: the object simulator, and the imaging setup. The former is designed for convenient generation of various objects with blinking emitters. We replace the projection lens of a commercial projector (Acer X113PH) by a microscope objective (40x, numerical aperture: NA=0.65) to de-magnify the projector pixels to $1.34 \times 1.34$ µm$^2$ squares at the object plane. Two irises, one placed in front of the projector and the other at the object plane, are used to block all the stray light generated by the projector. Light from the object passing through both scattering media and the imaging iris, is captured by a camera sensor (Andor Neo 5.5, 2560×2160 pixels, and 6.5-µm pixel size). The scattering media are a ground glass diffuser (a static one) or fresh chicken eggshell membrane (a dynamic one) in our demonstration. An optical filter (Thorlabs FB550-10, 550 nm wavelength, and 10 nm full-width at half-maximum - FWHM) is mounted on the camera to narrow the optical spectrum. Blinking emitters are generated by randomly blinking projector pixels. For invasive measurement of the point spread function (PSF), only one center pixel is turned on.

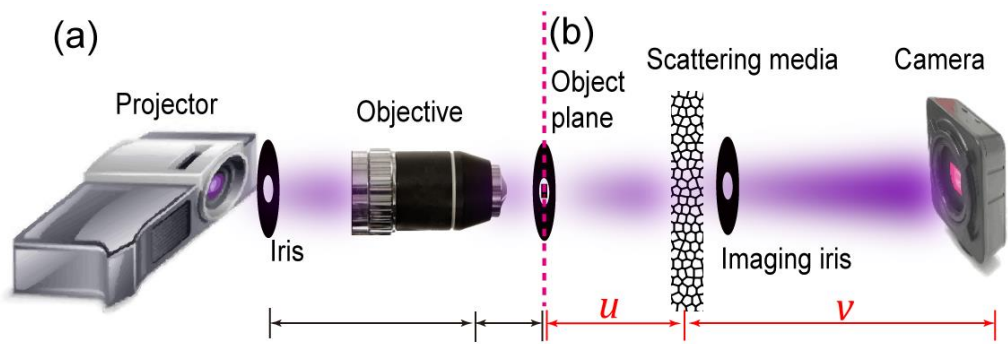

**Supplementary Fig. S1. Optical setup to demonstrate SOSLI for non-invasive super-resolution imaging through strongly scattering media. (a)** Object-simulator, which is designed for generating various microscopic objects at the object plane. **(b)** Simple optical configuration for imaging setup where u = 10 mm and v = 100 mm.

## The state-of-the-art non-invasive and invasive imaging

We conduct experiments for the demonstration of the current state-of-the-art non-invasive and invasive imaging through a 120-grit ground glass diffuser using the experimental setup shown in supplementary Fig. S1. A non-invasive image is retrieved from the autocorrelation of a single-shot speckle pattern by applying the phase retrieval algorithm. An invasive image is the deconvolution of a single-shot speckle pattern with an invasively measured PSF. The diameters of the imaging iris are set as 1 mm, 2 mm and 3 mm that correspond to NAs of 0.05, 0.1 and 0.15, respectively, and the diffraction limit of 6.7 µm, 3.4 µm and 2.3 µm respectively. We can easily see the effects of NA on resolution from the results given in supplementary Fig. S2. An imaging system is a low pass filter, where higher NA (higher cut-off frequency) provides higher resolution (i.e. sharper) images than lower NA does. Because of the effects of noise, camera's dynamic range and dark counts on the performance of the phase retrieval algorithm, the single-shot non-invasive images have a slightly lower resolution than the diffraction limit. If these effects are too high, the algorithm may not even converge. On the other hand, the deconvolution images have a slightly higher resolution than the diffraction limit because deconvolution recovers and enhances the high frequency components of images, i.e. sharper cut-off low pass filter.

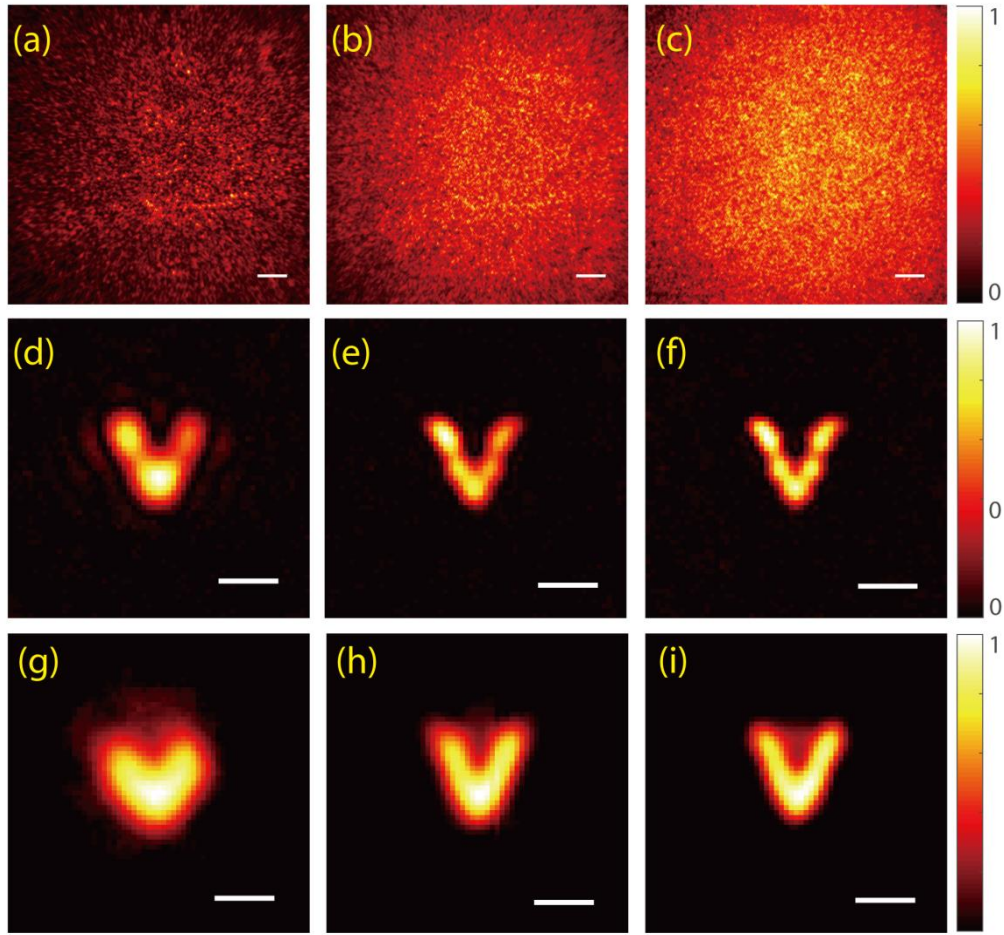

**Supplementary Fig. S2. The effect of NA on resolution for invasive and non-invasive imaging through scattering media. (a-c)** The speckle patterns of the same object with NAs of 0.05, 0.1, and 0.15, respectively. Experimental results of the state-of-the-art invasive **(d-f)** and non-invasive **(g)-(i)** imaging through the 120-grit ground glass diffuser with the speckle patterns in (a)-(c) respectively. Scale bar: 10 camera pixels, i.e. 65 µm on the imaging plane (see method).

# Estimating a point spreading function from a randomly selected stochastic emitter pattern

From a series of stochastic speckle patterns recorded for SOSLI, we pick up one pattern randomly (Fig. S3a). The iterative phase retrieval algorithm is utilized to retrieve the emitter pattern at low resolution as presented in Fig. S3b-c. Two similar emitter patterns shifted from each other can be retrieved from a single speckle pattern by two different runs of the algorithm because autocorrelation only keeps the relative emitter positions while losing their exact positions. Figure S3d-e present the emitter positions after localization. The localized-emitter images show very clean emitters, removing all the noise or artifacts of the phase retrieval algorithm. Figure S3f-g show the estimated PSFs calculated from a single speckle pattern (Fig. S3a) and two different phase retrieval/localization results (Fig. S3d-e). Figure S3i-j present the emitter positions localized after deconvolution of another speckle pattern (Fig. S3h) with two estimated PSFs in Fig. S3f-g. The emitter positions in Fig. S3i-j automatically align very well with those in Fig. S3d-e, respectively because they are reconstructed from the same estimated PSFs. The absolute positions of emitters and PSF are not important in our SOSLI. They do not affect our results and usual imaging techniques do not concern about absolute position.

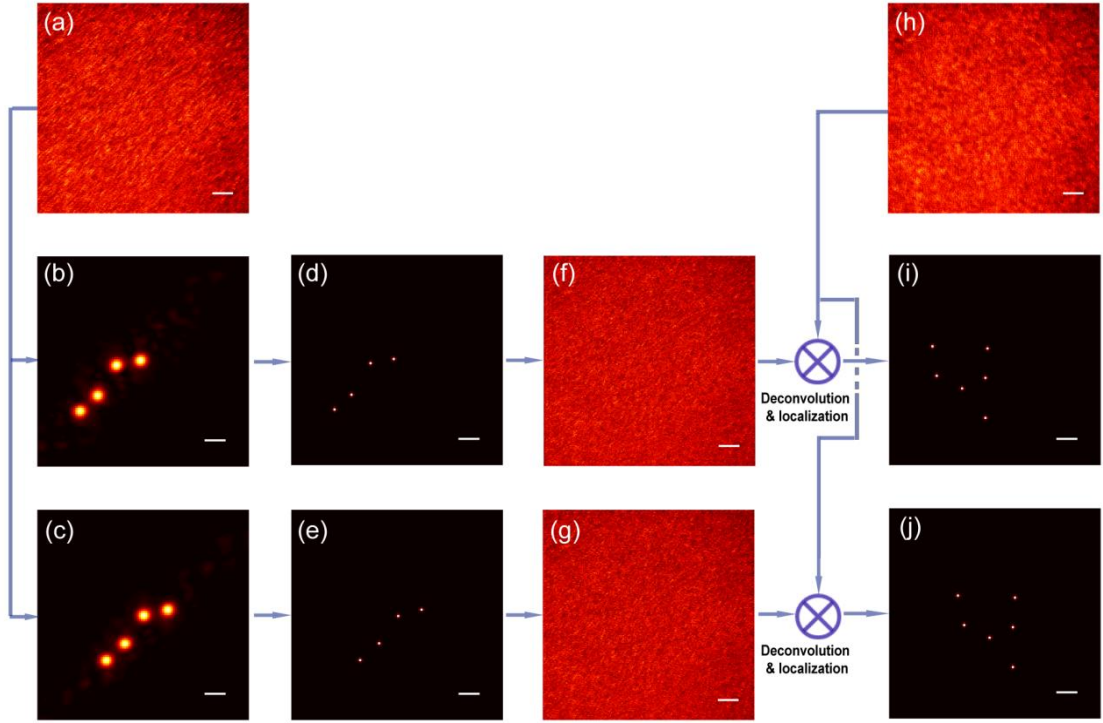

**Supplementary Fig. S3. Reconstructed and individual stochastic emitter pattern and estimated PSF for deconvolution. (a)** A typical stochastic speckle pattern. **(b-c)** Typical images retrieved from autocorrelation by phase retrieval algorithm. **(d-e)** Localized-emitter images. **(f-g)** Estimated PSFs. **(h)** Another stochastic speckle pattern. **(i-k)** Emitter positions calculated from a single speckle pattern and two estimated PSFs. Scale bar: 200 camera pixels for a, h, g, h and 10 camera pixels for b, c, d, e, i, j (see method).

## Reconstruction with PSFs estimated from different speckle patterns

During our implementation of SOSLI, we randomly choose one pattern out of multiple collected speckle patterns for estimation of the PSF. Interestingly, different speckle patterns give us slightly different estimated PSFs. The differences are not only arbitrary shift from each other but also the pattern itself (Supplementary Fig. S4a-c and S5b-c). However, the reconstructed images from these estimated PSFs are very much similar (Supplementary Fig. S4 d-f) with different shifts. It seems that we only estimate the main features of the PSF, which is sufficient for reconstruction. We test the similarity of these three estimated PSFs by calculating the correlation among them. The autocorrelation of PSF1 shows its random speckle nature with a bright spot (Gaussian profile) at the center (Supplementary Fig. S5a). The correlation of PSF1 with the other PSFs indicates an off-center bright spot with some background (Supplementary Fig. S5b-c), implying that these PSFs share the main features and shift from each other. The clear bright spot allows us to locate its center then obtain the relative shift between PSFs. The obtained relative shifts (Supplementary Fig. S5d) are exactly equal to the relative shifts between the retrieved objects (Supplementary Fig. S4 d-f).

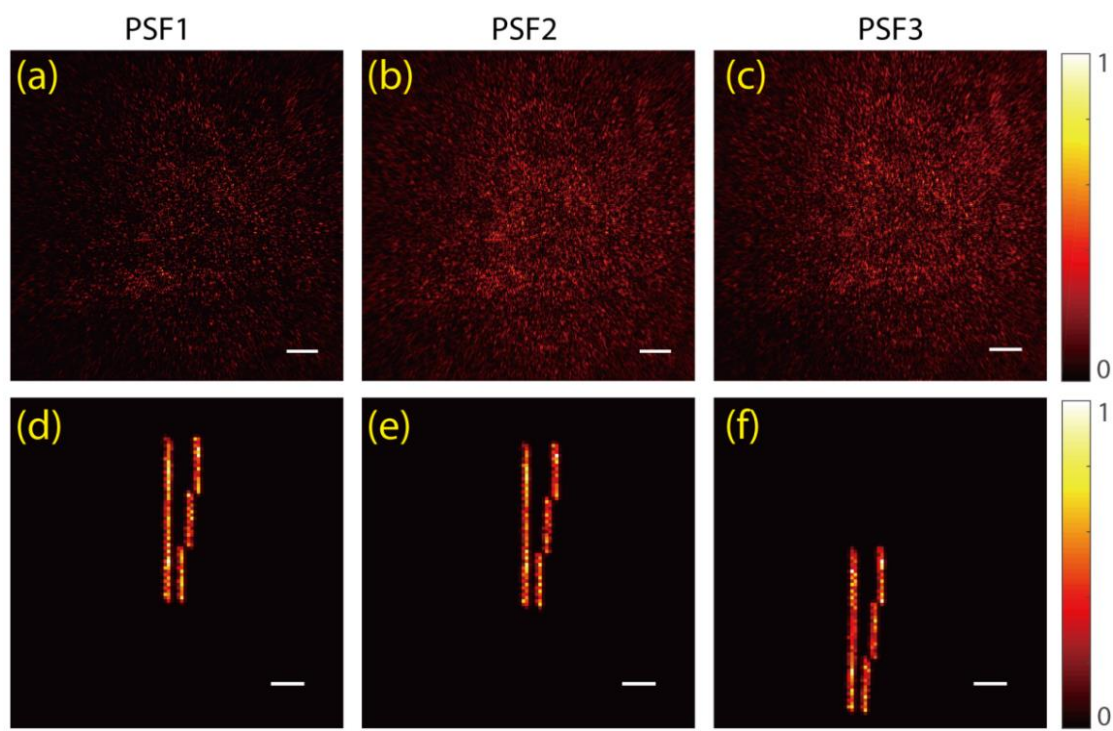

**Supplementary Fig. S4. Retrieved results of an object with the PSFs estimated from different speckle patterns arbitrarily picked. (a-c)** The estimated PSFs, **(d-f)** the corresponding reconstructed images at super-resolution. Scale bar: 200 camera pixels for a-c, and 10 camera pixels for d-f (see method).

The observation is interesting, important and useful. Because of the localization, we can easily remove the background artifact and noise in the phase retrieval images and deconvolution images. Therefore, the SOSLI approach can tolerate more error in PSF estimation. This not only explains how and why the SOSLI should work very well in static scattering media, but also inspires us to conceive a successful solution for dynamic scattering media in the section 7.

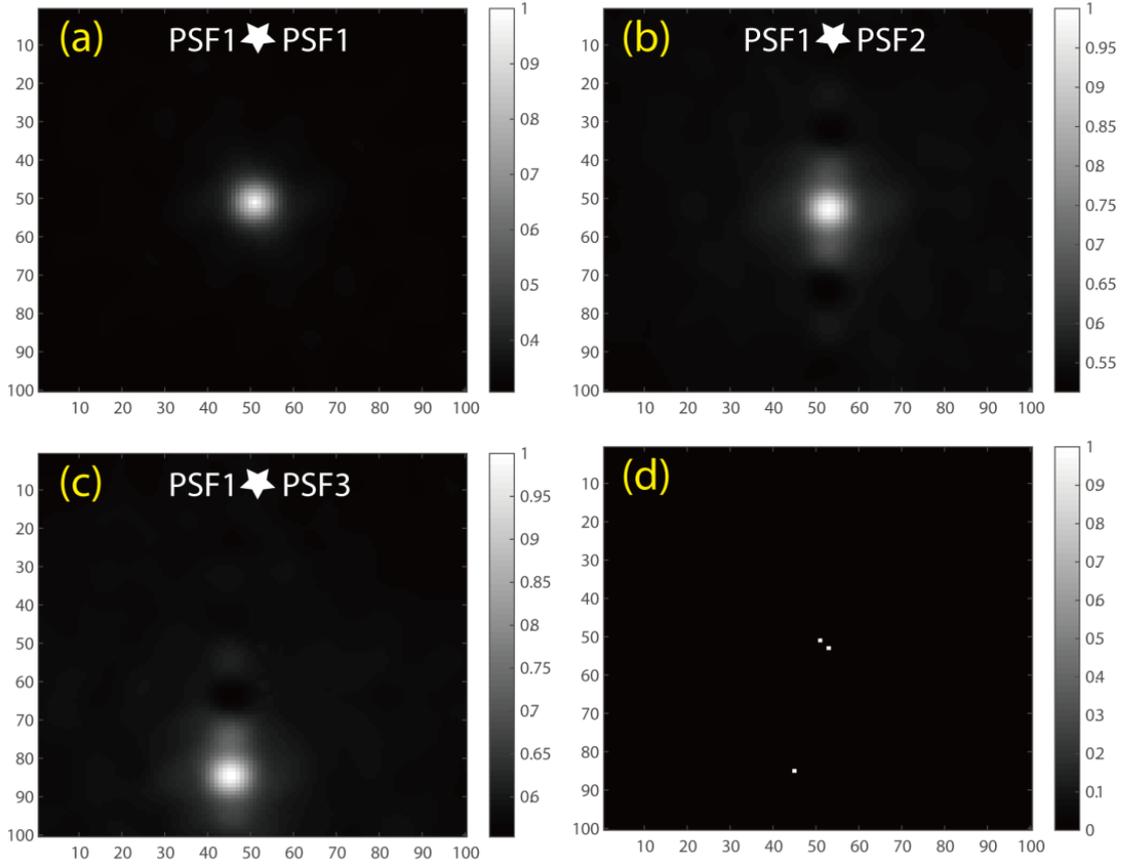

**Supplementary Fig. S5. Cross-correlations of the PSF1 with the other PSFs to obtain the relative shifts between them. (a-c)** The correlation patterns. **(d)** The relative shifts between PSF1 with other PSFs. The numbers on the side of image indicates the pixel numbers.

# Reconstruction of super-resolution images with different numbers of stochastic patterns.

In our experiment, the object to be imaged through a diffuser constitutes multiple blinking emitters. SOSLI technique reconstructs a super-resolution image from multiple stochastic patterns. The quality of the reconstructed image will increase with the number of frames. We characterize the reconstructed images with various numbers (n=200, 400, 800, and 8000) of the randomly blinking emitter patterns which are used Fig. 3c. The results are shown in supplementary Fig. S6 where supplementary Fig. S6e-h show the results of supplementary Fig. S6a-d after the bicubic interpolation processing. It is obvious that reconstruction with 8000 stochastic patterns gives the best image quality (supplementary Fig. S6d&h). However, 300-400 frames are reasonably good for the reconstruction of our simple object (Supplementary Fig. S6b&f).

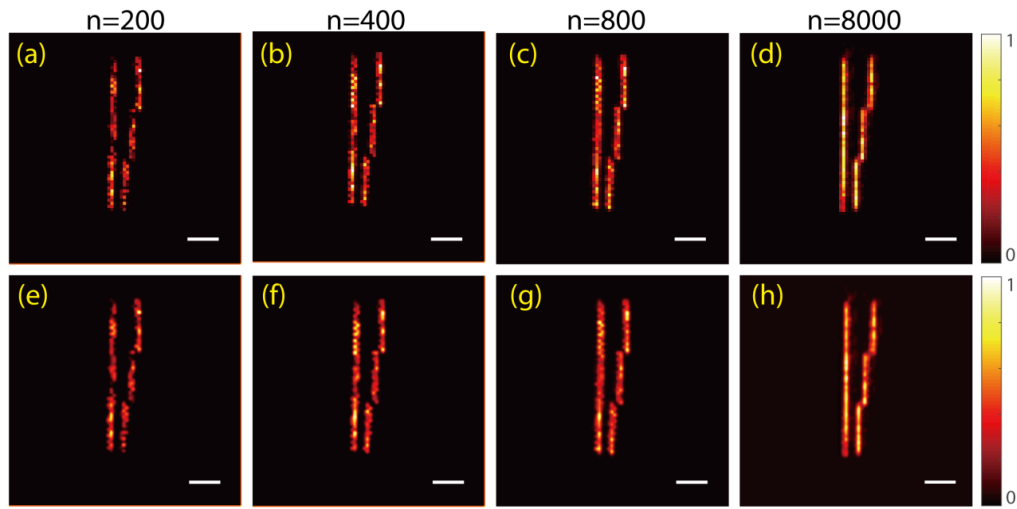

**Supplementary Fig. S6. Retrieved results of an object by SOSLI with the different numbers of stochastic patterns**. **(a-d)** The raw results from SOSLI. **(e-h)** The results after bicubic interpolation processing. Scale bar: 10 camera pixels, i.e. 65 µm on imaging plane (see method).

## Deconvolution results with a decorrelated PSF

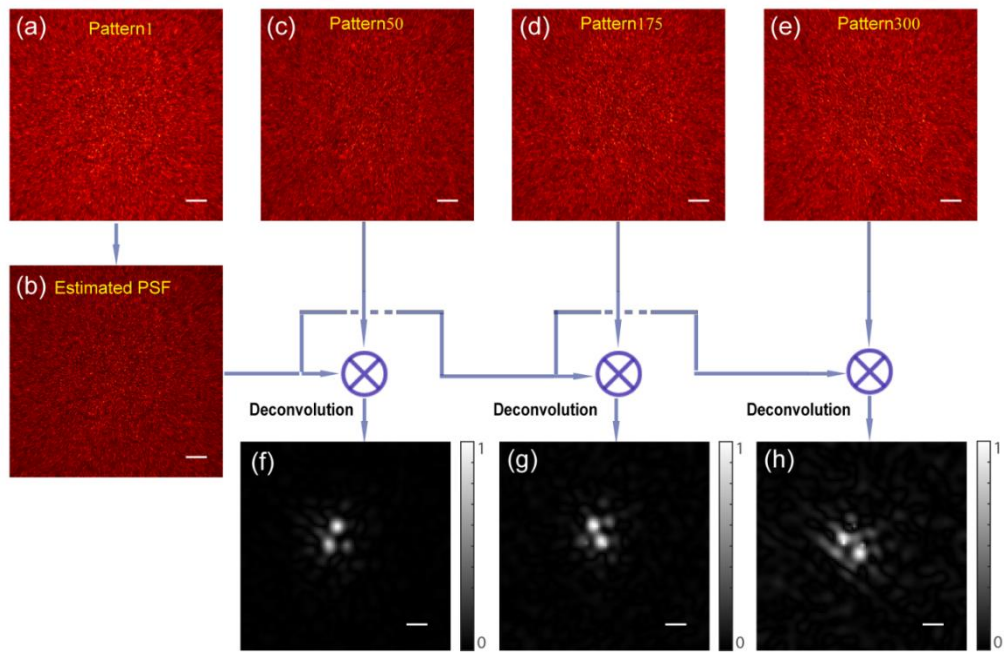

**Supplementary Fig. S7. Deconvolution results with a decorrelated PSF of chicken eggshell membrane. a)** The first stochastic speckle pattern is chosen to estimate the PSF. **b)** The estimated PSF from the first stochastic speckle pattern. **c-e)** The stochastic speckle patterns with the membrane's decorrelation (the correlation coefficients with respect to the pattern number are presented in Fig. 5a: 0.65, 0.36, 0.2 respectively). **f-h)** The deconvolution results of the speckle patterns in c-e, respectively, with the estimated PSF from the first speckle pattern.

Scale bar: 200 camera pixels for a-e, and 10 camera pixels for f-h (see method).

# Adaptive approach for SOSLI to do super-resolution imaging through a biological tissue

Our adaptive approach for SOSLI need to define the number of stochastic frames for each section depending on the decorrelation time and image acquisition time. The requirement is that every two adjacent sections have some correlation. In our approach, we even divide the set of patterns into overlapping sections (supplementary Fig. S8b) to guarantee correlation between adjacent sections. By doing sectional reconstruction, we effectively operate SOSLI with high speckle correlation for dynamic scattering media (shading area in Fig. 5a and supplementary Fig. S8a). Individual super-resolution images are reconstructed for each section independently. Before superposing these images on top of each other, we need to align them.

As presented in the previous section, the relative shifts between the reconstructed images are equal to the relative shifts between the estimated PSFs when the scattering media is not completely decorrelated. We calculate the correlation pattern of the estimated PSFs for every two adjacent sections to identify the relative shifts between them. The vectors indicated the relative shifts between section $i$ and section $j$ as follows.

$$\overrightarrow{P_{i,j}} = position\{PSF_i \star PSF_j\} - position\{PSF_i \star PSF_i\}$$

where *position*{ } gives 2D coordinate of the brightest spot center of an image, and $\star$ indicates the correlation calculation. $position\{PSF_i \star PSF_i\}$ is simply the center of the image because of speckle-type PSF. Supplementary Fig. S9a-j show the 10 vectors $\overrightarrow{P_{i,j}}$ for the frame set from 2 to 11 with respect to the first frame set. There relative positions are presented in supplementary Fig. S9k-l. With this, all the individual super-resolution images can be inversely shifted before superposing (supplementary Fig. S8c) to achieve the full image as presented in Fig. 5c.

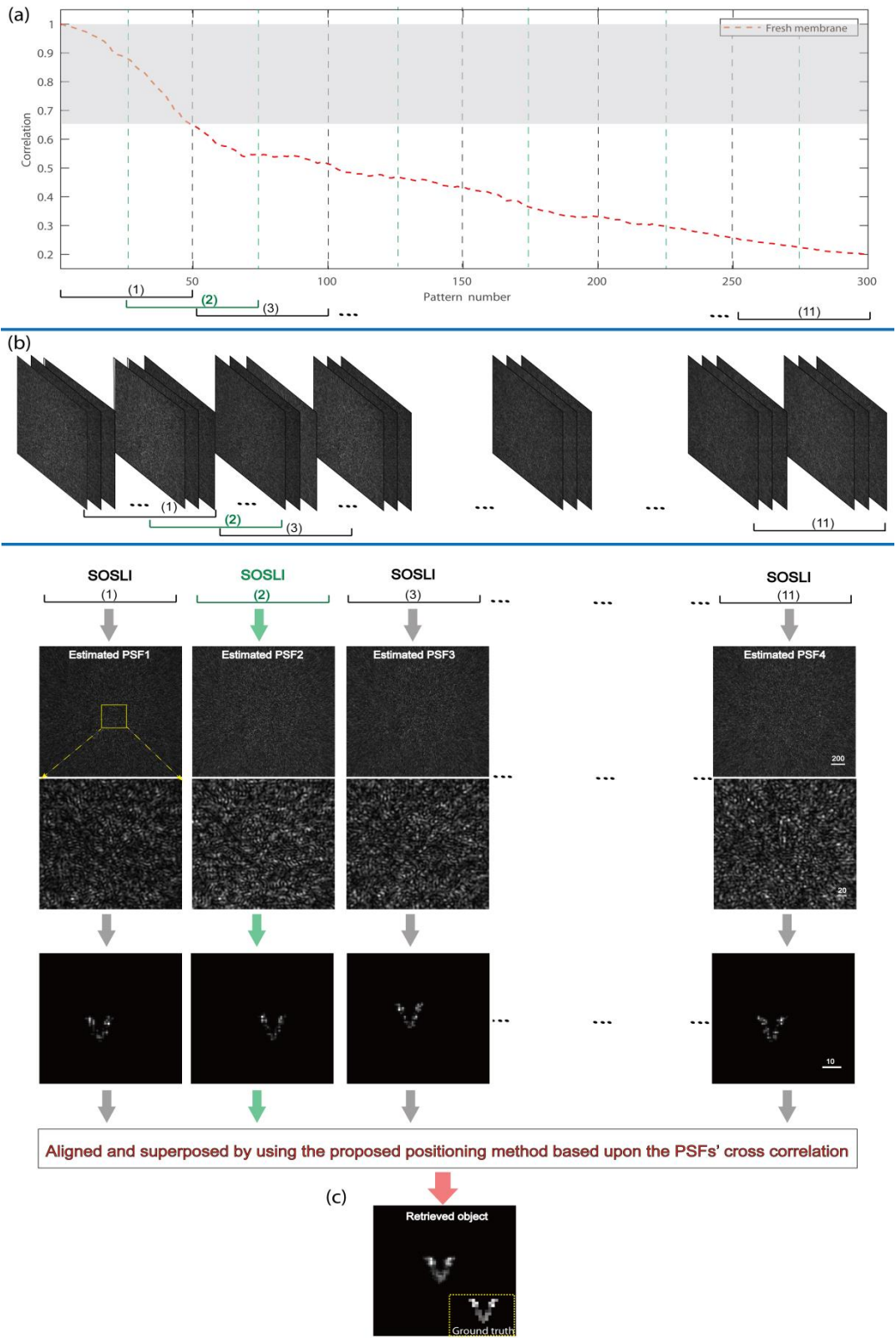

**Supplementary Fig. S8. Schematic for the implementation of SOSLI with adaptive PSFs for imaging through a biological tissue. (a)** The measured decorrelation characteristic of a chicken eggshell membrane. **(b)** Sectioning and overlapping of the collected speckle patterns. **(c)** Implementation of SOSLI with adaptive PSFs for retrieving the hidden object at super-

resolution. Scale bars: 200, 20 and 10 on the scale bar are camera pixels (see method).

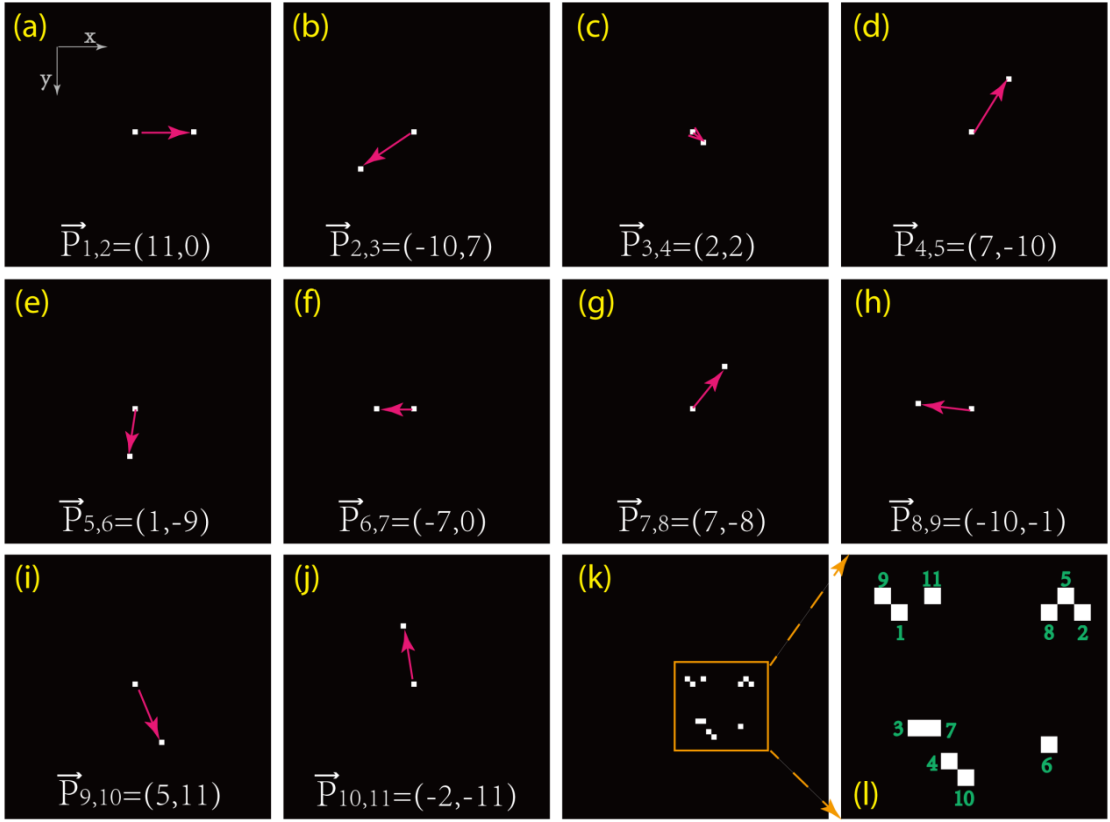

**Supplementary Fig. S9. The shifted vectors and the relative position of sectional super-resolution images. (a-j)** the shifted vectors $\vec{P_{I,J}}$ between two adjacent sections. **(k)** The relative positions of the 10 sectional reconstructed images (from 2 to 11) with respect to the first one. **(l)** Magnified center region of (k).

# The most adaptive approach for SOSLI for highly dynamic scattering media

By using the most adaptive SOSLI approach, we can align the emitter patterns even with decorrelation of two consecutive speckle pattern up to 80% (Fig. S10). The important point is to find the center of the brightest spot in correlation between two PSFs estimated from two consecutive speckle patterns. It might look similar to localizing the emitters from noisy deconvolution image due to 80% decorrelation in Fig. S7h. However, finding the brightest spot (there is only one) then localizing its center in Fig. S10h is relatively simpler than localizing all the centers of bright spots in Fig. S7h while we do not know the number of emitters.

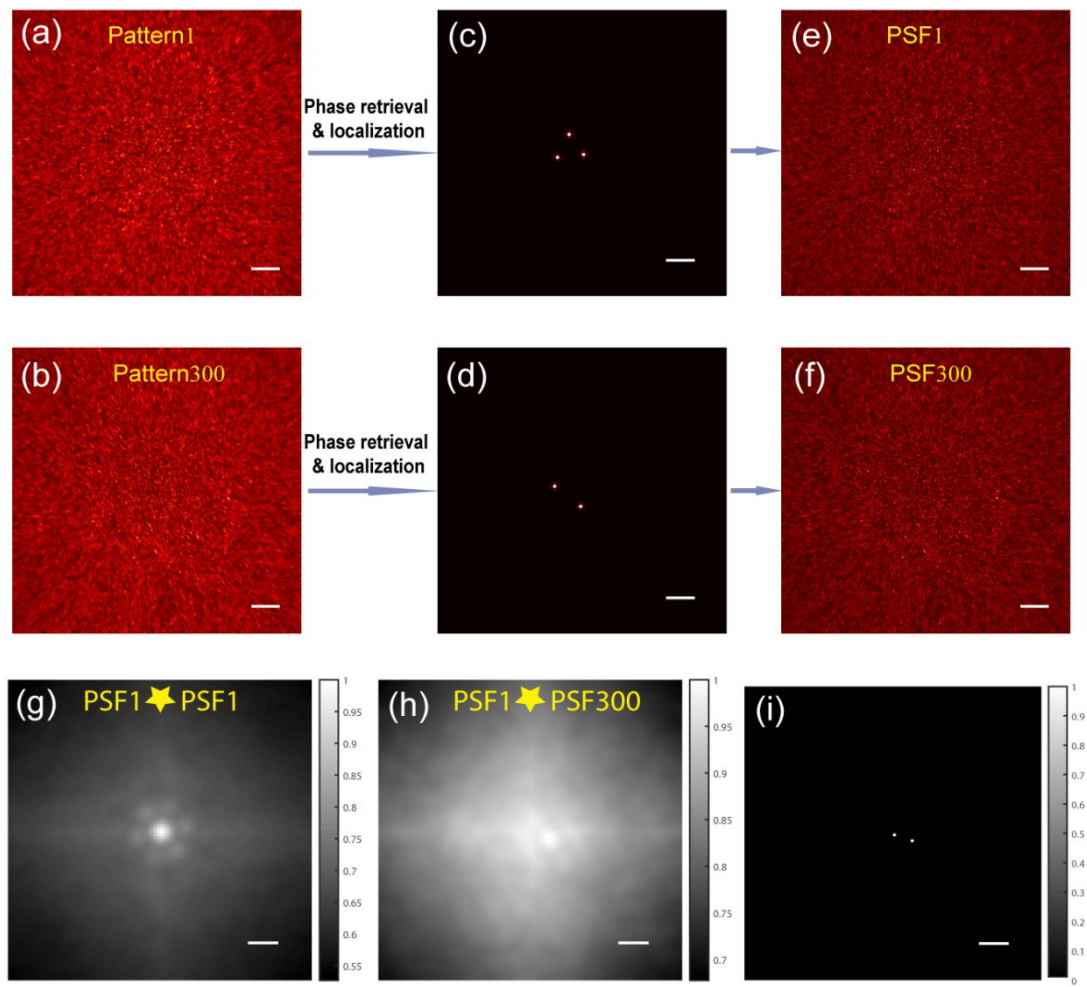

**Supplementary Fig. S10. The most adaptive approach for SOSLI to mitigate highly dynamic scattering media. a-b)** Two speckle patterns taken when scattering media decorrelate 80%, i.e. the correlation of 20%. **c-d)** Emitter patterns recovered from speckle patterns in a&b, respectively, by the phase retrieval algorithm and localization. **e-f)** The estimated PSFs from speckle patterns in a&b, respectively. **g)** The autocorrelation of PSF1 in figure e. **h)** The correlation between PSF1 (in figure e) and PSF300 (in figure f). **i)** The centers of brightest spots in g&h, showing the relative shift between two PSFs, which is also the relative shift between two emitter patterns in c&d. Scale bar: 200 camera pixels for a-b,e-f, and 10 camera pixels for c-d,g-i (see method).